\documentclass[a4paper,12pt]{article}

\usepackage{amsmath,amssymb,epsfig}

\newcommand{\beq}{\begin{equation}}
\newcommand{\eeq}{\end{equation}}
\newcommand{\ba}{\begin{array}}
\newcommand{\ea}{\end{array}}
\newcommand{\bea}{\begin{eqnarray}}
\newcommand{\eea}{\end{eqnarray}}
\newcommand{\bean}{\begin{eqnarray*}}
\newcommand{\eean}{\end{eqnarray*}}
\newcommand{\eref}[1]{(\ref{#1})}

\newcommand{\comment}[1]{}

\newcommand{\cM}{{\cal M}}
\newcommand{\cN}{{\cal N}}

\newcommand{\CN}{{\cal N}}

\newcommand{\cC}{{\cal C}}
\newcommand{\cX}{{\cal X}}
\newcommand{\IP}{\mathbb{P}}
\newcommand{\IC}{\mathbb{C}}

\newcommand{\IZ}{\mathbb{Z}}
\newcommand{\f}{{\cal F}^{\flat}}
\newcommand{\firr}[1]{{}^{{\rm Irr}}\!{\cal F}^{\flat}_{#1}}

\newcommand{\setall}{\setcounter{equation}{0}
        \setcounter{theorem}{0}}

\begin{document}
\begin{flushright}
Bicocca-FT-08-03\\
CERN-PH-TH/2008-001;
SISSA 02/2008/EP\\
Imperial/TP/08/AH/02;
NI08001\\ 
\end{flushright}
\vskip 0.25in

\renewcommand{\thefootnote}{\fnsymbol{footnote}}
\centerline{{\Huge Mastering the Master Space}}
~\\
{\bf
Davide Forcella${}^{1}$\footnote{\tt forcella@sissa.it}, 
Amihay Hanany${}^{2}$\footnote{\tt a.hanany@imperial.ac.uk, 
hanany@physics.technion.ac.il}, 
Yang-Hui He${}^{3}$\footnote{\tt hey@maths.ox.ac.uk}, 
Alberto Zaffaroni${}^{4}$\footnote{\tt alberto.zaffaroni@mib.infn.it}
}
~\\
~\\
{\hspace{-1in}
\scriptsize
\begin{tabular}{ll}
  ${}^1$ &{\it International School for Advanced Studies (SISSA/ISAS) \& INFN-
Sezione di Trieste, via Beirut 2, I-34014, Trieste, Italy} \\
  &{\it PH-TH Division, CERN CH-1211 Geneva 23, Switzerland}\\
  ${}^2$ 
  &{\it Department of Physics, Technion, Israel Institute of Technology,
    Haifa 32000, Israel}\\
  &{\it Theoretical Physics Group, Blackett Laboratory, 
    Imperial College, London SW7 2AZ, U.K.}\\
  ${}^3$
  & {\it Collegium Mertonense in Academia Oxoniensis, Oxford, OX1 4JD, U.K.}\\
  & {\it Mathematical Institute, Oxford University, 24-29 St.\ Giles', Oxford, 
OX1 3LB, U.K.}\\
  & {\it Rudolf Peierls Centre for Theoretical Physics, Oxford University, 1 
Keble Road, OX1 3NP, U.K.}\\
  ${}^4$
  & {\it Universit\`{a} di Milano-Bicocca and INFN, sezione di Milano-Bicocca, 
Piazza della Scienza, 3; I-20126 Milano, Italy}
\end{tabular}
}
\vspace{1cm}

\begin{abstract}
Supersymmetric gauge theories have an important but perhaps under-appreciated notion of
a master space, which controls the full moduli space. For world-volume theories of D-branes
probing a Calabi-Yau singularity $\cX$ the situation is particularly illustrative. 
In the case of one physical brane, the master space $\f$ is the space of F-terms and 
a particular quotient thereof is $\cX$ itself.
We study various properties of $\f$ 
which encode such physical quantities as Higgsing, BPS spectra, hidden global 
symmetries, etc. Using the plethystic program we also discuss what happens at 
higher number $N$ of branes. This letter is a summary and some extensions 
of the key points of a longer companion paper hep-th/0801.1585. 
\end{abstract}

\setcounter{footnote}{0}
\renewcommand{\thefootnote}{\arabic{footnote}}

\newpage

\section{Introduction}\setall
The vacuum of a quantum field theory is of vital physical significance.
For $\cN=1$ supersymmetric gauge theories in four dimensions, the vacuum is obtained by the space 
of solutions $D^\flat$ of {\bf D-terms}, coming from the gauge and matter 
content and the solution space $\f$ of {\bf F-terms}, coming from the critical 
points of the superpotential. 
This vacuum moduli space $\cM$ is typically a high dimensional object of subtle 
structure and consists of many branches, such as mesonic versus baryonic, and 
Higgs versus Coulomb, etc.
Conceptually, $\cM$ is a quotient of $\f$ by the gauge symmetries 
prescribed by $D^\flat$. In this short summary of a companion paper \cite{bigpaper} we would like to emphasize the role played by $\f$ and present it as a critical object in the study of supersymmetric gauge theories.

The study of $\f$ for a generic supersymmetric gauge theory is an important and a long term project. Here, as a starter, we focus on a special class of supersymmetric gauge theories where, in the  context of string theory, the $\cN=1$ gauge theory arises as the four
dimensional world-volume theory of a stack of $N$ coincident D3-branes 
transverse to a Calabi-Yau threefold singularity $\cX$. We look at the spectrum of chiral (BPS) operators in such theories and divide them into two types of gauge invariant operators which are typically called mesons (trace invariants) and baryons (determinant invariants). Correspondingly there are two types of moduli spaces, mesonic and baryonic, respectively, along which operators of the corresponding type admit a vacuum expectation value. The mesonic branch is referred to in the literature as the $N$-th symmetrized product\footnote{ 
Cf.~\cite{Berenstein:2002ge} for a consistency analysis of this identification.} of $\cX$. We do not focus on this point but give more attention on aspects of the baryonic branch and how it combines with the mesonic moduli space into a bigger space.

For $N=1$, a single D3-brane, the situation is particularly interesting: the 
mesonic branch is simply $\cX$; there are no gauge groups in the IR and $\cM 
\simeq \f$. 
$\f$ is called, in accord with the standard mathematical parlance, the {\bf 
master space} since its quotient is a moduli space \cite{master} and turns out to have some remarkable properties.
This forms a convenient starting point.
Moreover, for simplicity, $\cX$ is taken to be toric so that at least three $U(1)$ isometries are at hand.
For $N > 1$, a great deal can be learned via the plethystic program 
\cite{pleth,Butti:2006au,Forcella:2007wk,Butti:2007jv,Forcella:2007ps} despite the increasing subtlety in the structure of $\cM$.

This letter summarizes the key results of \cite{bigpaper} and uses the language of (computational) algebraic geometry (cf.~\cite{Gray:2006jb}); hence it outlines the requisite terminology where necessary.

\section{Theme and Variations in $\f$}
For $\cX$ an affine toric Calabi-Yau threefold and a single D3-brane, the gauge theory is a $U(1)^g$ quiver theory with $g$ nodes, $E$ bi-fundamental 
fields and a number $V$ of terms in the superpotential\footnote{Recently it is realised that the most conducive way of thinking of toric quiver gauge theories is via the language of {\bf dimer models/brane tilings} 
\cite{dimer}. This graphical method combines the matter content (quiver diagram) and the interaction terms (superpotential) into a single object: a periodic tiling of the 2-dimensional plane. It follows, for example, that
$V - E + g = 0$.}.
Decoupling the Abelian factors in the IR, we are left with the space of 
F-flatness $\f$ which is the principal object of our investigation.
The first two important properties of $\f$ are
\begin{enumerate}
\item 
{\em $\f$ is a toric variety of complex dimension $g+2$.}
This is so because as mentioned in the introduction,
$\cX \simeq \f // U(1)^{g}$, and an overall $U(1)$ decouples; thus 
$3 = \dim(\f) - (g-1)$. It is toric since it is acted upon by exactly $g+2$ $\IC^*$-actions corresponding to the classical global symmetries of the gauge theory: one R and two flavor, coming from the isometries of the toric threefold $\cX$, as well as $g-1$ baryonic, IR relic symmetries of the non-trivial $U(1)$ factors, some of which are anomalous. Specifically, we can define $\f$ as an affine algebraic variety in $\IC[x_1, \ldots, x_E]$ with appropriate $U(1)$ charges (weights) to the variables $x_i$ under the
$g+2$ dimensional toric action.
\item
The moduli space of gauge theories is well-known to have many branches;
this is reflected by the fact that {\em $\f$ is typically a reducible algebraic variety}. Either directly or using methods of toric ideals, we can perform {\bf primary decomposition} \cite{m2} thereupon to extract the irreducible pieces.
We find that it contains a top-dimensional irreducible component
of the same dimension and degree, 
as well as many smaller dimensional irreducible linear 
pieces, realised as coordinate hyperplanes. 
The top component is usually 
dubbed the {\bf coherent component}, which we 
denote as $\firr{}$. An interesting aspect of it is being a Calabi-Yau manifold of dimension $g+2$.
\end{enumerate}

Now, one of the most fundamental quantities which characterizes an algebraic 
variety $X$ is the {\bf Hilbert series}, which is the generating function for 
the number $\dim(X_i)$ of independent polynomials at a given degree on $X$:
\begin{equation}\label{hilb}
H(t;~X) = \sum_{i=0}^\infty \dim(X_i) t^i 
=
(1-t)^{-\dim(X)} P(t) \ .
\end{equation}
In the above, 
$P(t)$ 
turns out to be a polynomial with integer 
coefficients. We can readily refine \eref{hilb} by having a list $\underline
{t}$ of dummy variables in which case the number of polynomials of a given 
multi-degree would be counted. We can conveniently use \cite{m2} to compute the Hilbert series, or, alternatively, use the Molien formula as given below in \eref{molien}. 

The Hilbert series turns out to be of fundamental importance to supersymmetric gauge theories.
Physically, the Hilbert series is the key to the {\bf Plethystic program} 
\cite{pleth,Forcella:2007wk,Butti:2007jv}: the dummy variables are naturally identified with  {\em chemical potentials} associated to the multi-degrees which are combinations of $U(1)$-charges and the object of counting is the spectrum of mesonic and baryonic BPS operators. In \eref{hilb}, for example, $t$ can be taken to be the chemical potential\footnote{To be precise $t$ is identified with the ``fugacity" of the R-charge and $w=-\log t$ is identified as the chemical potential but we will call $t$ the chemical potential by abuse of notation.} for the R-charge. Plethystic exponentiation of the refined 
Hilbert series then counts all chiral BPS operators.

Upon studying $\f$ for a wealth of illustrative examples, we discover that the master space $\f$ and especially its coherent component $\firr{}$ enjoy many remarkable properties \cite{bigpaper}:

\begin{enumerate}
\item {\bf Symplectic quotient description of $\firr{}$: }
The traditional approach in understanding the moduli space of toric gauge theories is to use Witten's gauged linear sigma model (GLSM) where the spacetime fields are parameterized in terms of the latter \cite{toric}. We find an algebraic parameterization of the space of solutions of F-terms in terms of $c$ new fields $p_\alpha$, as $x_i =\prod_{\alpha=1}^c p_\alpha^{P_{i \alpha}}$ where $P_{i \alpha}$ turns out to be a matrix with 0 and 1 entries. The fields $p_\alpha$ are 
charged under  $c-g-2$ $U(1)$ gauge groups.
\comment{We have translated the space of solution of the algebraic equations defining $\f$ into the moduli space of an Abelian linear sigma model. 
}
In mathematical language, this is the symplectic quotient description of the {\em coherent component} of our affine toric variety
\begin{equation}\label{quotient}
\firr{~} = \mathbb{C}^c//(\IC^*)^{c-g-2} \ ,
\end{equation}
with the action specified by a $c$ by $c-g-2$ matrix of charges $Q$.
\comment{The number of fields $c$ and the matrix $Q$ are difficult to find in the traditional approach.} The GLSM fields can be associated with points in the 
toric diagram of $\cX$ with multiplicities \cite{toric}.

The GLSM description
allows us to compute the Hilbert series using a localisation formula based on the Molien integral, which projects onto $(\IC^*)^{c-g-2}$ invariants \cite{Butti:2007jv}:
\begin{equation}\label{molien}
H(\underline{t}; \firr{~} ) = 
\oint_{|z_i|=1} \prod_{i=1}^{c-g-2} \frac{dz_i}{z_i} \prod_{\alpha=1}^c \frac
{1}{1- y_\alpha \; {\underline z}^{{\underline q}^\alpha}} \ ,
\end{equation}
where ${\underline q}^\alpha$ is the vector of $U(1)^{c-g-2}$ charges of the fields $p_\alpha$, given by the $\alpha$-th column of $Q$, and $y_\alpha\equiv y_\alpha(\underline{t})$ is a monomial in $\underline{t}$ specifying the $U(1)$ global charges of $p_\alpha$.

\item{{\bf Dimer Model: }}
The GLSM fields $p_\alpha$ are now elegantly understood to be perfect matchings in the associated dimer model \cite{dimer} and the charge-matrix $Q$ to be given by the linear relations among these perfect matchings \cite{bigpaper}. We illustrate with the example for $\cX=dP_0 \equiv \IC^3/\IZ_3$ as given in
Figure \ref{f:pmdP0}.
\begin{figure}[t]
\begin{center}
\includegraphics[scale=0.4]{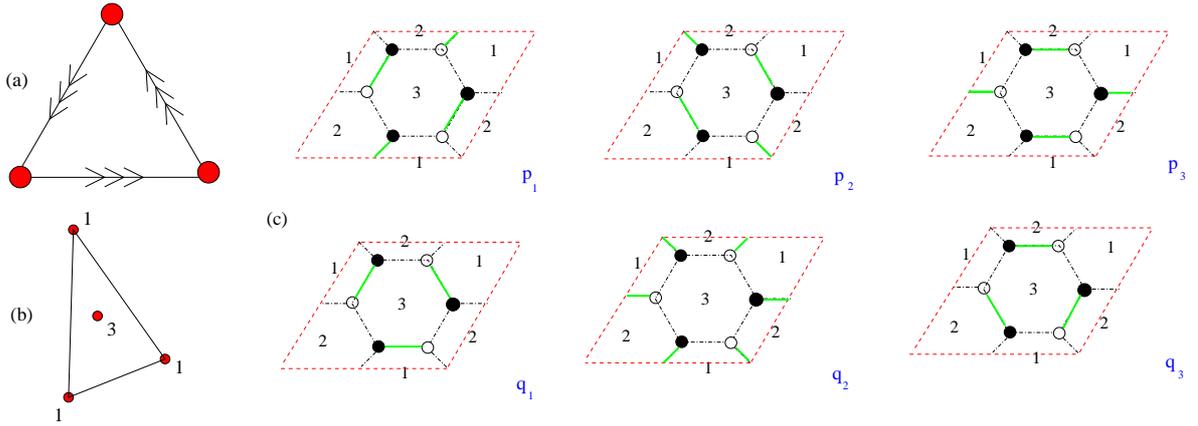} 
\caption{{\sf 
(a) The quiver diagram for $dP_0\equiv \IC^3/\IZ_3$ with gauge group $U(N)^3$ and nine fields with superpotential $W=\epsilon_{\alpha\beta\gamma} X^{(\alpha)}_{12} X^{(\beta)}_{23} X^{(\gamma)}_{31}$; \,
(b) The toric diagram, with the labeled multiplicity of GLSM fields; \, 
(c) The perfect matchings for the corresponding dimer model, with $p_i$ the external matchings and $q_i$, the internal. There is only one linear relation $-p_1-p_2-p_3+q_1+q_2+q_3=0$, giving the GLSM description of $\firr{\IC^3/\IZ_3}$ as $\mathbb{C}^6//[-1,-1,-1,1,1,1]$.}}
\label{f:pmdP0}
\end{center}
\end{figure}
In accord with the dimer language, 
we see that perfect matching generate the coherent 
component\comment{We thank Alastair King and Nathan Broomhead for first pointing this out to us. The precise relation is a consequence of the Birkhoff-von Neumann theorem investigated by their upcoming publication.}.

\item {\bf Surgery: }
The Hilbert series for the various irreducible 
pieces in the primary decomposition of $\f$ obey surgery rules \cite
{Hanany:2006uc} according to the intersection of the pieces. 

\item {\bf Calabi-Yau property: } 
The coherent component $\firr{}$ is affine Calabi-Yau\footnote{Recall that the symplectic quotient
\eref{quotient} is Calabi-Yau iff the vector of charges, given by the columns of $Q$, each sums to zero.}  
of dimension $g+2$. This easily follows from the dimer description as shown in \cite{bigpaper}.

\item {\bf Palindromic Hilbert series: }  
An intriguing property of the Hilbert Series for $\firr{~}$ is its symmetry.  
The numerator $P(t)$ of $H(t;\firr{})$, which we 
recall to be an integer polynomial of degree, say, $n$, has a palindromic 
symmetry for its coefficients $a_{j = 0, \ldots, n}$, viz., $P(t)$ is invariant under the exchange $a_j \leftrightarrow a_{n-j}$. This is a consequence of the 
Stanley theorem \cite{stanley} and the fact $\firr{}$ is toric Calabi-Yau.

\item {\bf Invariance under Seiberg duality: } 
We can perform the decomposition analysis to various toric/Seiberg dual phases for the same geometry \cite{Feng:2001bn}. We find a suggestive property and conjecture that for Seiberg dual phases, the coherent components of the master space are isomorphic as expected from the fact that Seiberg dual theories should have the same moduli space. 
However, the linear components differ, suggesting that some of the smaller-dimension pieces may be lifted by quantum corrections.

\item {\bf Linear Components and Flows: }
Physically, we can interpret the coherent component $\firr{}$ as the Higgs 
branch and the linear components, as the Coulomb branch of the moduli space. 
An archetypal theory which exhibits the branch structure of moduli spaces
is the $\IC^2 / \IZ_2$ orbifold theory, which has $\cN=2$ supersymmetry.
We find that the acquisition of vacuum expectation values (VEV) of the fields 
parameterizing the linear components induces flows in the gauge theory. 

Indeed, a gauge theory coming from a toric singularity $\cX$ can be Higgsed to another; in the toric diagram, this is seen as the deletion of nodes \cite{toric,Feng:2002fv}, or, geometrically, as the partial resolution of $\cX$. Many examples of this phenomen are demonstrated in \cite{bigpaper}.

The affine cone $F_0$ 
over the zeroth Hirzebruch surface,
for example, has a toric diagram which contains that of $\IC^2 / \IZ_2$. We
find that, upon primary decomposition of the master space $\f_{F_0}$ the
linear pieces are coordinate hyperplanes and giving VEV's to the variables
therein gives $\IC^2 / \IZ_2$. We can also find chains of toric theories  
flowing by the successive acquisition
of VEV's (i.e., deletion of nodes in the toric diagram) of fields
parameterizing the linear components of the master spaces, e.g.,
$dP_3 \rightarrow dP_2 \rightarrow F_0
\rightarrow \IC^2 / \IZ_2$, where $dP_n$ is the affine cone over the 
$n$-th del Pezzo surface.

\item {\bf Hidden Global Symmetries: } 
The moduli space of a field theory may possess symmetries beyond 
gauge or explicit global symmetries, which develop as the theory flows to the IR. We call them {\bf hidden global symmetries} since they are not
manifest in the UV Lagrangian. 

For D-brane gauge theories, the
UV symmetries of the Lagrangian are generically Abelian:
the three isometries of $\cX$ are visible in the UV as flavor symmetries. The Abelian gauge factors become weakly coupled in the IR and give rise to (possibly anomalous) baryonic symmetries. In \cite{Franco:2004rt}, additional symmetry structure of $\cX$ was investigated by grouping the fields according to representations of non-Abelian groups and the terms in the superpotential to invariants.

Such hidden global symmetries are surprisingly manifest in the
Hilbert series of the master space. 
As shown in the final Table, the full symmetry of $\firr{}$ is in many cases
non-Abelian.
The lesson we learn is that
{\em the terms in the refined Hilbert series of $\f$ 
can be arranged according to the representations of the hidden global symmetry
of the gauge theory}.

Let us illustrate with $dP_0=\IC^3/\IZ_3$. 
The quiver is shown in Figure \ref{f:pmdP0}. The master space is irreducible and given by $\f\simeq\firr{}=\mathbb{C}^6//[-1,-1,-1,1,1,1]$. We immediately see, by grouping the three $+1$ and three $-1$ in the charge matrix, that the symmetry of $\f$ is  $U(1)_R \times SU(3)_M\times SU(3)_H$. Here, $U(1)_R \times SU(3)_M$ is an obvious (mesonic) symmetry of the Lagrangian and corresponds to the symmetry of $\IC^3/\IZ_3$. On the other hand, $SU(3)_H$ is a hidden symmetry enhancing the two anomalous baryonic symmetries.

As re-writing of the refined Hilbert series reveals
that it organizes and decomposes according to the representation of the full
group $U(1)_R \times SU(3)_M\times SU(3)_H$:
\beq\label{c3z3}\ba{rcl}
g_1(t;\IC^3/\IZ_3) &=& \left(1 - [0,1,1,0] t^2 + ( [1,1,0,0] + [0,0,1,1] ) t^3 
 - [1,0,0,1] t^4 + t^6 \right) PE \left [ [1,0,0,1] t \right ] \\
& =& \sum\limits_{n=0}^\infty \left [n, 0, 0, n \right ] t^{ n } \ ,
\ea\eeq
where $[m,n,p,q]$ denotes the character of the product representation with 
weights $[m,n]$ under $SU(3)_M$ and $[p,q]$ under $SU(3)_H$. The nine
fundamental fields, whose plethystic exponential we are taking in \eref{c3z3}, transform as $[1,0,0,1]$. 

Therefore, we see that the hidden global symmetry of the theory is encoded
in the generating function which counts the BPS spectrum,
viz., the refined Hilbert series organizes according to the representations of the global symmetry. This phenomenon
persists for a host of illustrative and non-trivial geometries \cite{bigpaper}.
In particular, for $dP_{n = 1, \ldots, 8}$, the symmetry is the exceptional Lie group $E_n$  as conjectured in \cite{Franco:2004rt}. 
We present some of the results in the table at the end of this letter.

It is an interesting question to understand when the symmetry for one brane
extends to a general number $N$ of branes and thus to a hidden symmetry of the theory for some or all values of $N$. This is shown in a set of examples in \cite{bigpaper}.
The structure of the master space for an arbitrary number of branes also becomes subtle: The moduli space is given as a quotient of the space of F-terms by the non-Abelian factors of the gauge group since the Abelian $U(1)$ factors decouple in the IR. A further quotient by $U(1)^{g-1}$ leads to the mesonic moduli space which for $N$-branes is conjectured to be the symmetrized product of the Calabi-Yau singularity $\cX$. The master space thus has dimension $3N + g - 1$.

One of the salient features of the Plethystic program is that it addresses many properties of arbitrary number of branes with ease and without explicit knowledge of the actual space: the generating function for one brane $g_1(t)=H(t;\firr{})$ determines the generating function for arbitrary $N$. One can explictly check
that the $SU(4)_H$ symmetry for the conifold persist only up to $N=2$ (since
it mixes mesonic and baryonic symmetries and it enters in conflict with
the plethystic exponential which is performed in the sector with definite baryonic charge) while the hidden symmetries of $\IC^3/\IZ_3$ and $F_0$ extend to
arbitrary $N$.   
We summarize some of the above discussions with Table \ref{f:tab}.

\begin{table}[!!!]
$\begin{array}{|c|c|c|c|c|}  \hline
\cX & \dim(\f) & \firr{} & H(t;~\firr{}) & \mbox{Global Symmetry} \\ 
\hline \hline
\IC^3 & 3 & \IC^3 & (1-t)^{-3} & U(3) \\ \hline
\cC & 4 & \IC^4 & (1-t)^{-4} & U(1)_R \times SU(4)_H \\ \hline
(\IC^2 / \IZ_2) \times \IC & 4 & \cC \times \IC & \frac{1 + t}{(1-t)^4} &
  U(1)_R \times SU(2)_R \times U(1)_B \times SU(2)_H \\ \hline
\IC^3 / \IZ_2\times \IZ_2 & 6 & - & \frac{1+6 t+6 t^2+t^3}{(1-t)^6} &
  U(1)_R \times U(1)^2 \times SU(2)^3_H \\ \hline
SPP & 5 & \cC \times \IC^2 & \frac{1 + t}{(1-t)^5} &
  U(1)_R \times U(1)_M \times SU(2)_H^3 \\ \hline
dP_0 & 5 & \simeq \f & \frac {1+4t+t^2}{(1-t)^5} &
  U(1)_R \times SU(3)_M \times SU(3)_H \\ \hline
F_0 & 6 & \cC \times \cC & \frac{(1 + t)^2}{(1-t)^6} &
  U(1)_R \times U(1)_B \times SU(2)^2_M \times SU(2)_H^2 \\ \hline
dP_1 & 6 & -& \frac{1 + 4 t + 7 t^2 + 4 t^3 + t^4}{(1 - t)^6(1+t)^2} &
  U(1)_R \times SU(2)_M \times U(1)^3 \times SU(2)_H \\ \hline
dP_2 & 7 & -& \frac{1 + 2t + 5 t^2 + 2t^3 + t^4}{(1-t)^7(1+t)^2} &
  U(1)_R \times SU(2)_H  \times U(1)^5 \\ \hline
dP_3 & 8 & -& \frac{1 + 4t^2 + t^4}{(1 - t)^8(1+t)^2} &
  (SU(2) \times SU(3))_H \times U(1)^5 \\ \hline
\end{array}$
\caption{{\small The toric Calabi-Yau threefold $\cX$ is exemplified by the list in the left-most column, where $\cC$ is the conifold, SPP, the suspended pinched points, $F_0$, the cone over $\IP^1 \times \IP^1$ and $dP_n$ the cone over the $n$-th del Pezzo surface. For these we tabulate the dimension of the single-brane master space $\f$. The top component thereof, $\firr{}$ is always of the same dimension and is Calabi-Yau; we present, where possible, what this space is explicitly, as well
as its Hilbert series. We also record the global symmetry of the respective theories: the subscript $R$ denotes R-symmetry, $M$ denotes the symmetry of the mesonic branch, $B$ is the baryonic charge, and $H$ denotes the hidden global symmetry. Note that the rank of the global symmetry group is equal to the dimension of $\f$.}}
\label{f:tab}
\end{table}

\end{enumerate}

\comment{
\section*{Acknowledgements}
To Nathan Broomhead, Vadim Kaplunovsky, Alaistair King, Balazs Szendro\"i and David Tong we are much obliged for many enlightening discussions.
A.~H., Y.-H.~H., and A.~Z.~are indebted to the Newton
Institute in Cambridge for warm hospitality and support during part of this
work.
A.~H.~is grateful to the Perimeter Institute, the physics and mathematics departments at the University of Texas at Austin, and Stanford University where parts of this project were completed.
D.~F.~is supported in part by INFN and the Marie Curie
fellowship under the programme EUROTHEPHY-2007-1.
Y.-H.~H bows to the gracious patronage of Merton College, Oxford through the FitzJames Fellowship.
A.~Z.~ is supported
in part by INFN and MIUR under contract 2005-024045-004 and
2005-023102 and by the European Community's Human Potential Program
MRTN-CT-2004-005104.
}


\end{document}